\documentclass[12pt]{article}

\def\lsim{\mathrel{\rlap{\lower3pt\hbox{\hskip0pt$\sim$}}
     \raise1pt\hbox{$<$}}}         %less than or approx. symbol
\def\gsim{\mathrel{\rlap{\lower4pt\hbox{\hskip1pt$\sim$}}
     \raise1pt\hbox{$>$}}}         %greater than or approx. symbol

\usepackage{amsmath}
\usepackage{amsfonts}

\begin{document}
\begin{titlepage}
\begin{flushright}
CLNS 95/1380
\end{flushright}
\bigskip
\bigskip

\centerline{\Large \bf Asymmetric Orbifolds and Wilson Lines}
\medskip

\centerline{Zurab Kakushadze\footnote{\, Current address: Quantigic$^\circledR$ Solutions LLC, 1127 High Ridge Road \#135, Stamford, CT 06905 ({\it DISCLAIMER: This address is used by the corresponding author for no
purpose other than to indicate his professional affiliation as is customary in
publications. In particular, the contents of this paper
are not intended as an investment, legal, tax or any other such advice,
and in no way represent views of Quantigic® Solutions LLC,
the website \underline{www.quantigic.com} or any of their other affiliates}) and Department of Physics, University of Connecticut, 1 University Place, Stamford, CT 06901; email: zura@quantigic.com.} and S.-H. Henry Tye}

\bigskip

\centerline{Newman Laboratory of Nuclear Studies}
\centerline{Cornell University, Ithaca, NY 14853-5001, USA}

\medskip
\centerline{(December 19, 1995)\footnote{\, This paper appeared on arXiv on December 20, 1995, arXiv:hep-th/9512156.}}

\bigskip
\medskip

\begin{abstract}
{}We generalize the rules for the free fermionic string construction
to include other asymmetric orbifolds.
Examples are given to illustrate the use of these rules.
\end{abstract}
\end{titlepage}

\newpage
\section{Introduction}

{}The construction of string models has a long history. The number
of consistent string models is clearly very large (One may consider
various string models as different classical string vacua of a single
theory; in this case, we are talking about the construction of classical
vacua). The best understood string models are probably those obtained
via toroidal compactification, and also their orbifolds \cite{DHVW}.
However,
classification of such orbifold string models is still largely unexplored.
This is in part due to the lack of simple rules for constructing such models,
in particular, for asymmetric orbifolds \cite{NSV}.

{}The first class of asymmetric orbifold string models are the
free fermionic string models \cite{KLT}. Although this class of models
is rather restrictive (allowing only multiple ${\bf Z}_2$ twists),
the rules for such model building are quite simple.
As a result, rather complicated models
can be readily constructed \cite{CGL}, sometimes with the help of
computers.

{}A general framework for asymmetric orbifolds was given in \cite{NSV},
and it provides a setting for all orbifold models, since
some symmetric orbifolds at fixed radii
may be considered as special cases of asymmetric orbifolds. However, this
general approach is not easy to use in actual practice; as a result,
asymmetric orbifold models are not well explored. In this paper, we
simplify the construction of asymmetric orbifold models by presenting
explicit and rather simple rules for their model buildings.

{}The rules for the asymmetric orbifold construction are
quite similar to those for the free fermionic string models. Here we shall
follow the notations of Ref. \cite{KLST}. We shall
impose the consistency requirements on the string one-loop partition
function in the light-cone gauge: ({\em i}) one-loop modular invariance;
({\em ii}) world-sheet supersymmetry (if present),
which insures space-time Lorentz invariance in the covariant gauge;
and ({\em iii}) physically
sensible projection; this means the contribution of a
space-time fermionic (bosonic) degree of freedom to the partition function
counts as minus (plus) one. In all cases that can be checked, this
condition plus the one-loop modular invariance and factorization imply
multi-loop modular invariance.

{}The building blocks for any specific string model partition function
are the
(appropriate) characters for world-sheet fermions and bosons. The fermion
characters are the same as the ones used in the free fermionic
string models. The characters for bosons are combined from two types: those
for twisted chiral bosons
and those for chiral lattices. A key to obtaining simple rules is the
choice of basis for the chiral lattices. They are chosen so that, up to
phases, all the chiral lattice characters are permuted under any modular
transformation, as is the case for the chiral fermion characters.
Our discussion shall focus on heterotic
strings compactified to four spacetime dimensions; the generalization to
other dimensions and to Type II strings is straightforward.

{}As in the free fermionic string model constructions, the rules may
be used to build new models without direct reference to their original
partition functions and/or the characters in them. This turns out to be
useful because the partition function can get rather complicated.
In this paper, we consider models with Wilson lines and only one twist.
The rules given here can be used as a basis for further generalization to
the non-Abelian orbifold case, which will be discussed elsewhere.
For a general lattice, the sublattice invariant under the twist may be
difficult to identify. Sometimes, it is easier to start with a lattice
whose invariant sublattice is obvious, and then introduce background fields,
in particular Wilson lines, that commute with the twist. In fact, this
approach is very useful in the symmetric orbifold construction \cite{INQ}.
Here, our work can be considered as a generalization of their work to
the asymmetric orbifold case.

{}In section \ref{secII}, we briefly review the fermion and the boson characters
that we shall use later. In section \ref{secIII}, we derive the
rules for model building. To be specific, we shall consider only heterotic
strings compactified to four spacetime dimensions. Also, we shall confine
ourselves to the elementary particle sectors, where the conformal field
theory description (as given by the characters mentioned above) is sufficient.
The rules for model building are summarized.
In section \ref{secIV}, some examples of asymmetric orbifold models, with and without
Wilson lines, are explicitly constructed to illustrate the
rules. In section \ref{secV}, we discuss a model with higher-level gauge group. For
the sake of completeness, we present a couple of
symmetric orbifolds as well, with and without Wilson lines, in section \ref{secVI}.
Section \ref{secVII} contains the discussion and remarks. Some of the details are
relegated to the appendices.

%============================================================================

%\newpage
\section{Preliminaries}\label{secII}
\bigskip

%==============================================================================
\subsection{Framework}
\medskip

{}In this subsection we set up the framework for the remainder of this paper.
To be specific, we consider heterotic strings compactified to four
space-time dimensions. In the light-cone gauge, which we adopt, we have
the following world-sheet degrees of freedom:
One complex boson $\phi^0$ (corresponding to two transverse
space-time coordinates); three right-moving complex bosons
$\phi^\ell_R$, $\ell=1,2,3$ (corresponding to six internal coordinates);
four right-moving complex fermions $\psi^r$, $r=0,1,2,3$
($\psi^0$ is the world-sheet superpartner of the right-moving component of
$\phi^0$, whereas $\psi^\ell$ are the world-sheet superpartners of
$\phi^\ell_R$, $\ell=1,2,3$);
eleven left-moving complex bosons $\phi^\ell_L$,
$\ell=4,5,...,14$ (corresponding to twenty-two internal coordinates).
Before orbifolding, the
corresponding string model has $N=4$ space-time supersymmetry and the
internal momenta span an even self-dual Lorentzian lattice $\Gamma^{6,22}$.

{}It is convenient to organize the string states into sectors labeled by the
monodromies of the string degrees of freedom. Thus, consider
the sector where
\begin{eqnarray}\label{group1}
 &&\psi^r ({\overline z} e^{-2\pi i} ) =\exp (-2\pi i V^r_i)
 \psi^r ({\overline z}) ~,\nonumber\\
 &&\phi^\ell_R ({\overline z} e^{-2\pi i} )=\exp (-2\pi i W^\ell_i)
 \phi^\ell_R ({\overline z}) -U^\ell_i~,~~~\ell=1,2,3 ~,\\
 &&\phi^\ell_L ({z} e^{2\pi i} )=\exp (-2\pi i W^\ell_i )
 \phi^\ell_L ({z}) +U^\ell_i ~,~~~\ell=4,...,14 \nonumber
\end{eqnarray}
(Note that $\phi^0 ({z} e^{2\pi i}, {\overline z} e^{-2\pi i} )=\phi^0 ({z},
{\overline z})$ since $\phi^0$ corresponds to space-time coordinates).
These monodromies can be combined into a single vector
\begin{equation}\label{V_i}
 V_i =(V^0_i (V^1_i ~~  (W^1_i ,U^1_i ))(V^2_i ~~  (W^2_i ,U^2_i ))
 (V^3_i ~~  (W^3_i ,U^3_i ))\vert\vert (W^4_i ,U^4_i)...
 (W^{14}_i, U^{14}_i )) ~.
\end{equation}
The double vertical line separates the right- and left-movers.
Without loss of generality we can restrict the values of $V^r_i$ and
$W^\ell_i$ as follows: $-{1\over 2} \leq
V^r_i < {1\over 2}$; $0\leq W^\ell_i <1$
(A complex boson (fermion) with boundary condition
$W^\ell_i~(V^r_i)=0$ or $1\over 2$ can be split into two real bosons
(fermions)). The shifts $U^\ell_i$ can be
combined into a real $(6,22)$ dimensional Lorentzian
vector ${\vec U}_i$ defined up to
the identification ${\vec U}_i \sim {\vec U}_i+{\vec P}$,
where ${\vec P}$ is an arbitrary vector of $\Gamma^{6,22}$.

{}The monodromies (\ref{group1}) can be viewed as fields $\Phi$ (where $\Phi$
is a collective notation for the fields $\psi^r$, $\phi^\ell_R$ and
$\phi^\ell_L$) being periodic $\Phi ({z} e^{2\pi i}, {\overline z}
e^{-2\pi i} )=\Phi ({z},{\overline z})$ up to the identification
$\Phi \sim g(V_i) \Phi g^{-1} (V_i)$, where $g(V_i)$ is an element of the
{\em orbifold} group $G$. In this paper we will only consider the cases where
$G$ is an {\em Abelian} group. For two elements $g(V_i)$ and $g(V_j)$
to commute, we must have
$U^\ell_i =0$ if $W^\ell_j \not=0$, and $U^\ell_j =0$ if $W^\ell_i \not=0$.

{}This leads us to a simpler form of
$V_i$ where instead of having
a double entry $(W^\ell_i ,U^\ell_i)$ for each complex boson
we will specify
either $W^\ell_i$ (whenever $W^\ell_i \not=0$, in which case $U^\ell_i =0$),
or $U^\ell_i$ (whenever $U^\ell_i \not=0$, in which case $W^\ell_i =0$).
To keep track of whether a given entry
corresponds to a twist or a shift, it is convenient to
introduce an {\em auxiliary} vector
\begin{equation}
 W=(0 (0~W^1) (0~W^2) (0~W^2)
 \vert\vert
 W^4 ~...~W^{14} )~.
\end{equation}
The entries $W^\ell$ are defined as follows: $W^\ell={1\over 2}$ if in at
least one sector (labeled by, say, $V_i$) of the model the corresponding
boson has twisted boundary conditions ({\em i.e.}, $W^\ell_i \not=0$); $W^\ell
=0$, otherwise. For example, if
\begin{equation}
 W=(0 (0~{1\over 2})^3 \vert\vert 0^{11})~,\nonumber
\end{equation}
then
\begin{equation}
 V_i =(V^0_i (V^1_i ~W^1_i)(V^2_i ~W^2_i)(V^3_i ~W^3_i)\vert\vert
 U^4_i ~...~U^{14}_i ) \nonumber
\end{equation}
is {\em a priori} compatible with $W$.
Here $W^1_i$, $W^2_i$ and $W^3_i$ correspond to the twists,
$U^4_i$,...,$U^{14}_i$ correspond to the shifts, and $V^r_i$, $r=0,1,2,3$,
specify the fermionic spin structures.

{}The notation we have introduced proves convenient in describing the
sectors of a given string model based on the orbifold group $G$. For $G$ to be
a finite discrete group, the element $g(V_i)$ must have a finite order $m_i
\in {\bf N}$,
{\em i.e.} $g^{m_i} (V_i)=1$. This implies that $V^r_i$ and $W^\ell_i$ must be
rational numbers, and the shift vector ${\vec U}_i$ must be a rational
multiple of a vector in $\Gamma^{6,22}$; that is, $m_i V^r_i , m_i W^\ell_i
\in {\bf Z}$, and $m_i {\vec U}_i \in \Gamma^{6,22}$. To describe all the
elements of the group $G$, it is convenient to introduce the set of
generating vectors $\{ V_i \}$ such that
${\overline {\alpha V}}={\bf 0}$ if and only if
$\alpha_i \equiv 0$. Here ${\bf 0}$ is the null vector:
\begin{equation}
 {\bf 0}=(0 (0~0)^3 \vert\vert 0^{11})~.
\end{equation}
Also, $\alpha V \equiv \sum_i \alpha_i V_i$
(The summation is defined as $(V_i +V_j )^\ell=V^\ell_i +
V^\ell_j$), $\alpha_i$ being integers that
take values from $0$ to $m_i -1$. The overbar notation is defined as follows:
${\overline {\alpha V}} \equiv \alpha V -\Delta(\alpha)$, and the components
of ${\overline {\alpha V}}$ satisfy $-{1\over 2}\leq {\overline {\alpha V}}
^r <{1\over 2}$, $0\leq {\overline {\alpha W}}^\ell<1$;
here $\Delta^r (\alpha),\Delta^\ell (\alpha)
\in {\bf Z}$. So the elements of the group $G$ are in one-to-one
correspondence with the vectors ${\overline {\alpha V}}$ and will be denoted
by $g({\overline {\alpha V}}$). It is precisely the Abelian nature of $G$ that
allows this correspondence (by simply taking all possible linear
combinations of the generating vectors $V_i$).

{}Now we can identify
the sectors of the model. They are labeled by the vectors
${\overline {\alpha V}}$, and in a given sector ${\overline {\alpha V}}$ the
monodromies of the string degrees of freedom are given by
$\Phi ({z} e^{2\pi i}, {\overline z}
e^{-2\pi i} )=g({\overline {\alpha V}}) \Phi (z,{\overline z}) g^{-1}
({\overline {\alpha V}})$.

{}$G$ is a symmetry of the
Hilbert space of the original string model with $N=4$ supersymmetry
compatible with the operator algebra of the underlying
(super) conformal field theory. If $\vert
\chi \rangle$ is a state in the original Hilbert space,
$g({\overline {\alpha V}})\vert \chi \rangle$
(where there must exist a representation of $g({\overline {\alpha V}})$ via
the vertex operators of the theory)
also belongs to the same Hilbert space. One
consequence of this requirement is that $G$ must commute with the
Virasoro algebra, which is indeed the
case for the class of Abelian asymmetric orbifolds considered in this paper.
We must also require that $G$ (anti)commutes with the right-moving
super-Virasoro algebra (which ensures space-time Lorentz invariance in the
covariant gauge).
This implies the following {\em supercurrent} constraint
\begin{equation}\label{supercurrent}
 V^\ell_i + W^\ell_i =V^0_i \equiv s_i ~(\mbox{mod}~1)~,~~~\ell=1,2,3~.
\end{equation}
Here $s_i$ is the monodromy of the supercurrent
${\overline S}({\overline z}e^{-2\pi i} )=\exp(2\pi is_i)
{\overline S}({\overline z})$, which must satisfy $s_i \in {1\over 2}{\bf Z}$.
Then the sectors with ${\overline {\alpha V}}^0 =0$
give rise to space-time bosons, while
the sectors with ${\overline {\alpha V}}^0
=-{1\over 2}$ give rise to space-time fermions.

%==============================================================================
\subsection{Fermion and Boson Characters}
\medskip

{}Let us confine our attention to the orbifolds
with a single twist of prime order, generated
by the $V_1$ vector
(The  order of this twist is defined as
the smallest positive integer $t_1$, such that $\forall\ell~t_1 W^\ell_1
\in {\bf Z}$; note that $t_1$ is a divisor of $m_1$).

{}In a given sector ${\overline {\alpha V}}$, the right- and left-moving
Hamiltonians are given by the corresponding sums of the Hamiltonians for
individual string degrees of freedom. The Hilbert space in the
${\overline {\alpha V}}$ sector
is given by the momentum states $\vert {\vec P}_
{\overline {\alpha V}} +\alpha {\vec U} \rangle$, and also
the states obtained from these states by acting with the fermion and boson
creation operators (oscillator excitations). In the untwisted sectors, that
is, sectors ${\overline {\alpha V}}$ with $\alpha_1 =0$, we have ${\vec P}_
{\overline {\alpha V}} \in \Gamma^{6,22}$. In the twisted sectors
${\overline {\alpha V}}$ with $\alpha_1 \not=0$, we have ${\vec P}_
{\overline {\alpha V}} \in
{\tilde I}$, where ${\tilde I}$ is the lattice dual to the lattice $I$, which
in turn is the sublattice of $\Gamma^{6,22}$ invariant under the action of
the twist part of the group element $g(V_1)$. This lattice must have a prime
$N_I$, where $N_I$ is the smallest positive integer such that
for all vectors ${\vec P}\in {\tilde I}$, $N_I {\vec P}^2\in 2{\bf Z}$;
moreover, for the corresponding characters to have the correct modular
transformation properties, it must be the case that either $N_I =1$ (in which
case $I$ is an even self-dual lattice), or $N_I =t_1$ (in which case $I$ is
even but not self-dual).

{}Now we turn to expressing the group elements $g({\overline {\beta V}})$
(in a given sector ${\overline {\alpha V}}$) in
terms of the generators of twists $J^\ell_{\overline {\alpha V}}$, shifts
${\vec P}_{\overline {\alpha V}}$, and $U(1)$ rotations of the
right-moving complex fermions $-N^\ell_{\overline {\alpha V}}$
(see Appendix \ref{appA}):
\begin{equation}
 g({\overline {\beta V}}) =
 \exp (2\pi i \beta V \cdot {\cal N}_{\overline {\alpha V}} +{1\over 2}
 \nu(\alpha_1 ,\beta_1) {\vec P}^2_{\overline {\alpha V}} )~.
\end{equation}
Here $\nu
(\alpha_1 ,\beta_1)$ is an integer taking value between $0$ and $N_I -1$
defined as
\begin{equation}
 \alpha_1 \nu(\alpha_1 ,\beta_1) =\beta_1 ~(\mbox{mod}~N_I)~,
 ~~~\alpha_1 \not=0~,
\end{equation}
and $\nu(0,\beta_1)\equiv 0$. ${\cal N}_{\overline {\alpha V}}=(
N^r_{\overline {\alpha V}},J^\ell_{\overline {\alpha V}},
{\vec P}_{\overline {\alpha V}})$, and the dot product is understood
with respect to the following signature:
\begin{equation}\label{dotproduct}
 \beta V \cdot {\cal N}_{\overline {\alpha V}} \equiv
  \beta{\vec U} \cdot {\vec P}_{\overline {\alpha V}} +
 \sum_{r} ({\beta V})^r N^r_{\overline {\alpha V}}
 \epsilon^r + \sum_\ell ({\beta W})^\ell J^\ell_{\overline {\alpha V}}
 \epsilon^\ell ~.
\end{equation}
The dot product of the vectors $\beta {\vec U}$ and ${\vec P}_
{\overline {\alpha V}}$ is
understood with respect to the Lorentzian metric
$\mbox{diag}((-)^6 , (+)^{22})$. The
signature $\epsilon^r$ for fermions equals $+1$ for
left-moving complex fermions, and $-1$ for
right-moving complex fermions, respectively. The
signature $\epsilon^\ell$ for bosons equals $-1$ for
left-moving complex bosons, and $+1$ for
right-moving complex bosons, respectively.

{}In section \ref{secIII} we express the
one-loop modular invariant partition function for an
orbifold model as a linear combination of the following characters:
\begin{equation}\label{totalchar}
 {\cal Z}^{\overline {\alpha V}}_{\overline {\beta V}} \equiv
 \mbox{Tr} ( q^{H^L_{\overline {\alpha V}}}
 ~{\overline q}^{H^R_{\overline {\alpha V}}} g^{-1} ({\overline {\beta V}})) ~.
\end{equation}
Here $H^L_{\overline {\alpha V}}$ and $H^R_{\overline {\alpha V}}$ are the
left- and right-moving Hamiltonians, respectively. The trace is taken over the
states in the Hilbert space corresponding to the sector
${\overline {\alpha V}}$. These characters can be computed as products of
building blocks, or contributions of individual string degrees of freedom,
which are reviewed in Appendix \ref{appA}. The result
can be written as a product of the corresponding fermion and boson
characters:
\begin{equation}
 {\cal Z}^{\overline {\alpha V}}_{\overline {\beta V}}=
 {\overline Z}^{\overline {\alpha V}}_{\overline {\beta V}}
 {\cal Y}^{\overline {\alpha V}}_{\overline {\beta V}} ~.
\end{equation}
The fermion characters
${\overline Z}^{\overline {\alpha V}}_{\overline {\beta V}}$ read:
\begin{equation}
 {\overline Z}^{\overline {\alpha V}}_{\overline {\beta V}}=
 \prod_{r} {\overline Z}^{{\overline {\alpha V}}^{\, r}}_
 {{\overline {\beta V}}^{\, r}}
\end{equation}
(The characters ${\overline Z}^v_u$ for a right-moving fermion are complex
conjugates of the characters $Z^v_u$ for a left-moving fermion given by
(\ref{fermionZ})).

{}The boson characters
${\cal Y}^{\overline {\alpha V}}_{\overline {\beta V}}$ read:
\begin{eqnarray}
 &&{\cal Y}^{\overline {\alpha V}}_{\overline {\beta V}}=
 Y^{\alpha {\vec U}}_{\beta {\vec U}} ~,~~~\alpha_1=\beta_1 =0~,\\
 &&{\cal Y}^{\overline {\alpha V}}_{\overline {\beta V}}=
 \xi (\alpha_1)
 Y^{\alpha_1 , \alpha{\vec U}}_{\beta_1 , \beta{\vec U}}
 \prod_{\ell=1}^3
 {\overline X}^
 {{\overline {\alpha W}}^\ell}_{{\overline {\beta W}}^\ell}
 \prod_{\ell=4}^{14}
 X^{{\overline {\alpha W}}^{\ell}}_{{\overline {\beta W}}^{\ell}}~,~~~
 \alpha_1 +\beta_1 \not=0~
\end{eqnarray}
(The characters ${\overline X}^v_u$ for a right-moving boson are complex
conjugates of the characters $X^v_u$ for a left-moving boson
given by (\ref{bosonX})).
The product over $\ell$ does {\em not} include terms with
${\overline {\alpha W}}^\ell ={\overline {\beta W}}^\ell =0$.

{}$Y^{\alpha {\vec U}}_{\beta {\vec U}}$ are the characters for the even
self-dual lattice $\Gamma^{6,22}$, whereas
$Y^{\alpha_1 , \alpha{\vec U}}_{\beta_1 , \beta{\vec U}}$ are the characters
for the lattice $I$ (If $I$ is an even self-dual lattice then instead of
$Y^{\alpha_1 , \alpha{\vec U}}_{\beta_1 , \beta{\vec U}}$ we would have to
use the characters similar to $Y^{\alpha {\vec U}}_{\beta {\vec U}}$ but
defined for the lattice $I\subset\Gamma^{6,22}$):
\begin{eqnarray}
 &&Y^{\alpha {\vec U}}_{\beta {\vec U}}=
 {1\over{\eta^{22} (q) {\overline \eta}^{6} ({\overline q}) }}
 \sum_{{\vec P}\in \Gamma^{6,22}}
 q^{{1\over 2}({\vec P}^L+\alpha{\vec U}^L )^2}
 {\overline q}^{{1\over 2}({\vec P}^R+\alpha{\vec U}^R )^2}
 \exp(-2\pi i\beta{\vec U} \cdot {\vec P}) ~,\\
 &&Y^{\alpha_1 , \alpha{\vec U}}_{\beta_1 , \beta{\vec U}} =
 {1\over{\eta^d (q) {\overline \eta}^{d^\prime} }}
 \sum_{{\vec P}\in {\tilde I}}
 q^{{1\over 2}({\vec P}^L+\alpha{\vec U}^L )^2}
 {\overline q}^{{1\over 2}({\vec P}^R+\alpha{\vec U}^R )^2} \times\nonumber\\
 &&\,\,\,\,\,\,\,\times\exp(-2\pi i(\beta{\vec U} \cdot {\vec P}+{1\over 2}\nu(\alpha_1 ,\beta_1)
 {\vec P}^2 )) ~.
\end{eqnarray}
Here $I$ has the Lorentzian metric $((-)^{d^\prime} ,(+)^d )$.
${\vec P}^L$,
${\vec P}^R$, and ${\vec U}^L$, ${\vec U}^R$, are the
left- and right-moving parts of the momentum and shift vectors, respectively.

{}The integers $\xi(\alpha_1)$
are nothing but the number of fixed points in the twisted sectors ($M_I$ is the
determinant of the metric of $I$) \cite{NSV}:
\begin{equation}
 \xi(\alpha_1) =M^{-{1\over 2}}_I
 \prod_{\ell} 2\sin(\pi {{\overline {\alpha W}}^\ell})
 ~,~~~\alpha_1 \not=0~,
\end{equation}
and $\xi(0)=1$ (The product over $\ell$ does {\em not} include terms with
${\overline {\alpha W}}^\ell =0$ ).

{}Under the $S$- and $T$-modular transformations the characters
${\cal Z}^{\overline {\alpha V}}_{\overline {\beta V}}$ transform as follows:
\begin{eqnarray}\label{calZ}
 &&{\cal Z}^{\overline {\alpha V}}_{\overline {\beta V}}
 \stackrel{S}{\rightarrow} \exp(2\pi i {\overline {\alpha V}}\cdot
 {\overline {\beta V}} )
 {\cal Z}^{\overline {\beta V}}_{\overline {-\alpha V}} ~,
 ~~~\alpha_1 \beta_1 =0~,\\
 &&{\cal Z}^{\overline {\alpha V}}_{\overline {\beta V}}
 \stackrel{S}{\rightarrow} \exp(2\pi i ({\overline {\alpha V}}-W)\cdot
 ({\overline {\beta V}}-W)+\chi (\alpha_1 ,\beta_1) )
 {\cal Z}^{\overline {\beta V}}_{\overline {-\alpha V}} ~,
 ~~\alpha_1 \beta_1 \not=0~,\\
 &&{\cal Z}^{\overline {\alpha V}}_{\overline {\beta V}}
 \stackrel{T}{\rightarrow} \exp(2\pi i({1\over 2}
 {\overline {\alpha V}} \cdot {\overline {\alpha V}}
 -{\overline {\alpha V}} \cdot W +{1\over 2})
 {\cal Z}^{\overline {\alpha V}}_{\overline {\beta V-\alpha V +V_0 }}~.
\end{eqnarray}
Here $V_0$ is the vector with $-1/2$ entry for each world -sheet fermion and
zero otherwise:
\begin{equation}
 V_0 =(-{1\over 2}(-{1\over 2}~0)^3 \vert\vert 0^{11})~.
\end{equation}
According to the above modular transformation properties of
${\cal Z}^{\overline {\alpha V}}_{\overline {\beta V}}$, $V_0$ is always among
the generating vectors $V_i$
(The sector corresponding to $V_0$ is the Ramond sector of the original
heterotic string). The dot product of two vectors
${\overline {\alpha V}}$ and ${\overline {\beta V}}$ is defined as in
(\ref{dotproduct}). For example,
\begin{equation}
 V_i \cdot V_j =
  {\vec U}_i \cdot {\vec U}_j +
 \sum_{r} V^r_i V^r_j
 \epsilon^r + \sum_\ell W^\ell_i W^\ell_j
 \epsilon^\ell ~.
\end{equation}

%============================================================================

%\newpage
\section{Orbifold Rules}\label{secIII}
\bigskip

{}In this section we derive the rules for constructing consistent string models
in the framework discussed in section \ref{secII}. The contribution to
the orbifold one-loop
partition function is a linear combination of the characters
${\cal Z}^{\overline {\alpha V}}_{\overline {\beta V}}$:
\begin{equation}\label{orbifold_partition_function}
 {\cal Z} = {1\over {\prod_{i} m_i }} \sum_{\alpha ,\beta}
 C^{\overline {\alpha V}}_{\overline {\beta V}}
 {\cal Z}^{\overline {\alpha V}}_{\overline {\beta V}} ~.
\end{equation}

{}The coefficients $C^{\overline {\alpha V}}_{\overline {\beta V}}$
must be such that (\ref{orbifold_partition_function}) is modular invariant.
Taking into account the modular transformation properties (\ref{calZ}), we have
the following constraints on the coefficients
$C^{\overline {\alpha V}}_{\overline {\beta V}}$ coming from the requirement
of modular invariance of (\ref{orbifold_partition_function}):
\begin{eqnarray}\label{Cmodinv}
 S:~~~&&C^{\overline {\alpha V}}_{\overline {\beta V}}
 \exp(2\pi i {\overline {\alpha V}}\cdot
 {\overline {\beta V}} ) =
 C^{\overline {\beta V}}_{\overline {-\alpha V}} ~,
 ~~~\alpha_1 \beta_1 =0~,\\
      &&C^{\overline {\alpha V}}_{\overline {\beta V}}
 \exp(2\pi i ({\overline {\alpha V}}-W)\cdot
 ({\overline {\beta V}}-W)+\chi (\alpha_1 ,\beta_1) ) =\nonumber\\
 &&\,\,\,\,\,\,\,C^{\overline {\beta V}}_{\overline {-\alpha V}} ~,
 ~~~\alpha_1 \beta_1 \not=0~,\\
 T:~~~&&C^{\overline {\alpha V}}_{\overline {\beta V}}
 \exp(2\pi i({1\over 2}
 {\overline {\alpha V}} \cdot {\overline {\alpha V}}
 -{\overline {\alpha V}} \cdot W +{1\over 2}) =
 C^{\overline {\alpha V}}_{\overline {\beta V-\alpha V +V_0 }}~.
\end{eqnarray}
In addition to (\ref{Cmodinv}) we require that for any physical sector labeled
by ${\overline {\alpha V}}$ the sum over $\beta$'s in
(\ref{orbifold_partition_function}) form a proper projection with eigenvalues
$0$ or $\xi (\alpha_1)$. Specifically, this means that
\begin{equation}\label{fourier}
 {1\over {\prod_i m_i}}
 \sum_{\beta} C^{\overline {\alpha V}}_{\overline {\beta V}}
 g^{-1} ({\overline {\beta V}}) = e^{2\pi i \alpha s} \eta(
 {\overline {\alpha V}}, {\cal N}_{\overline {\alpha V}}, {\vec P}^2_
 {\overline {\alpha V}}) ~,
\end{equation}
where $\eta({\overline {\alpha V}}, {\cal N}_{\overline {\alpha V}},
{\vec P}^2_{\overline {\alpha V}})$ takes values $0$ or $1$ depending on the
values of $\alpha_i$, ${\cal N}_{\overline {\alpha V}}$ and
${\vec P}^2_{\overline {\alpha V}}$. As a consequence, we have in this case
\begin{equation}
 {\cal Z}= \mbox{Tr} ( q^{H^L_{\overline {\alpha V}}}
 ~{\overline q}^{H^R_{\overline {\alpha V}}} \xi(\alpha_1) e^{2\pi i \alpha s}
 \eta({\overline {\alpha V}}, {\cal N}_{\overline {\alpha V}},
 {\vec P}^2_{\overline {\alpha V}}) )~.
\end{equation}
This is precisely the physically sensible projection; space-time bosons
contribute into the partition function with the weight plus one, whereas
space-time fermions contribute with the weight minus one (Each with degeneracy
$\xi(\alpha_1)$ due to fixed points in the twisted sectors).

{}The formal solution to (\ref{fourier}) is given by
\begin{equation}\label{C_coeffs}
 C^{\overline {\alpha V}}_{\overline {\beta V}} =
 \exp(2\pi i [\beta\phi
 ({\overline {\alpha V}})+{\alpha s} ])~.
\end{equation}

{}The phases $\phi_i ({\overline {\alpha V}})$ are constrained due to
(\ref{Cmodinv}):
\begin{eqnarray}\label{S1}
S:~&&\beta \phi({\overline {\alpha V}}) +\alpha \phi({\overline {\beta V}})
 +\alpha s +\beta s +{\overline {\alpha V}}\cdot
 {\overline {\beta V}} =0~(\mbox{mod}~1)~,~\alpha_1 \beta_1 =0~,\\
      &&\beta \phi({\overline {\alpha V}}) +\alpha \phi({\overline {\beta V}})
 +\alpha s +\beta s + \nonumber\\
\label{S2}
 &&\,\,\,\,\,\,\,({\overline {\alpha V}}-W)\cdot
 ({\overline {\beta V}}-W)+\chi (\alpha_1 ,\beta_1) =
 0~(\mbox{mod}~1)~,~\alpha_1 \beta_1 \not=0~,\\
\label{T}
 T:~&&\alpha\phi({\overline {\alpha V}}) +\phi_0 ({\overline {\alpha V}})
 +{1\over 2}{\overline {\alpha V}} \cdot {\overline {\alpha V}}
 -{\overline {\alpha V}} \cdot W +{1\over 2} =0~(\mbox{mod}~1)~.
\end{eqnarray}

{}Provided that $2 V_1 \cdot W \in {\bf Z}$, we have:
\begin{equation}
 \chi(\alpha_1, \beta_1) + W\cdot W -{\overline {\alpha V}} \cdot W-
 {\overline {\beta V}} \cdot W \equiv  0 ~({\mbox{mod}~ 1})~,
 ~~~\alpha_1 \beta_1 \not=0~,
\end{equation}
and the solution to the system of
equations (\ref{S1}), (\ref{S2}) and (\ref{T}) is given by:
\begin{equation}\label{phases}
 \phi_i ({\overline {\alpha V}}) =\sum_j
 k_{ij} \alpha_j + s_i -V_i \cdot {\overline {\alpha V}}
 ~(\mbox{mod}~1)~.
\end{equation}

{}The structure constants $k_{ij}$ must satisfy the following constraints:
\begin{eqnarray}
 \label{k1}
 &&k_{ij} +k_{ji} =V_i \cdot V_j ~({\mbox{mod}~ 1})~,\\
 \label{k2}
 &&k_{ij}m_j =0 ~({\mbox{mod}~ 1})~,\\
 \label{k3}
 &&k_{ii} +k_{i0} +s_i +V_i \cdot W
 -{1\over 2} V_i \cdot V_i =0 ~({\mbox{mod}~ 1})
\end{eqnarray}
(Note that there is {\em no} summation over repeated indices).

{}All the states are projected out of the sum in
(\ref{orbifold_partition_function}) except those satisfying
\begin{eqnarray}\label{sgf}
 &&V_i \cdot {\cal N}_{\overline {\alpha V}} =\phi_i ({\overline
 {\alpha V}})~(\mbox{mod}~1)~,~~~i\not=1~\mbox{or}~\alpha_1 =0~,\\
 &&\alpha_1 V_1 \cdot {\cal N}_{\overline {\alpha V}} +{1\over 2}
 {\vec P}^2_{\overline {\alpha V}}=\alpha_1 \phi_1 ({\overline
 {\alpha V}})~(\mbox{mod}~1)~,~~~\alpha_1 \not=0~.
\end{eqnarray}
This is the spectrum generating formula (Note that in the twisted sectors
($\alpha_1 \not= 0$) all the states appear with the multiplicity
$\xi (\alpha_1)$).

{}The states that satisfy the spectrum generating formula include
both on- and off-shell states.
The on-shell states must satisfy the additional constraint that
the left- and right-moving energies are equal. In the
${\overline {\alpha V}}$ sector they are given by:
\begin{eqnarray}\label{LRenergy}
 &&E^L_{\overline {\alpha V}} = -{1\over 2} +\nonumber\\
 &&\sum_{\ell:~\rm{\scriptstyle{left}}}
 \{ {1\over 2} {\overline {\alpha W}}^\ell (1-{\overline {\alpha W}}^\ell)+
 \sum_{q=1}^{\infty} [(q+{\overline {\alpha W}}^\ell -1)n^\ell_q +
 (q-{\overline {\alpha W}}^\ell ){\overline n}^\ell_q ]\} +\nonumber\\
 &&\sum_{q=1}^{\infty} q (n^0_q +{\overline n}^0_q )+{1\over 2}
 ({\vec P}^L_{\overline {\alpha V}}+{\vec U}^L)^2 ~,\\
 &&E^R_{\overline {\alpha V}} =-1 +\nonumber\\
 &&\sum_{\ell:~\rm{\scriptstyle{right}}}
 \{ {1\over 2} {\overline {\alpha W}}^\ell (1-{\overline {\alpha W}}^\ell)+
 \sum_{q=1}^{\infty} [(q+{\overline {\alpha W}}^\ell -1)m^\ell_q +
 (q-{\overline {\alpha W}}^\ell ){\overline m}^\ell_q ]\} +\nonumber\\
 &&\sum_{q=1}^{\infty} q (m^0_q +{\overline m}^0_q )+{1\over 2}
 ({\vec P}^R_{\overline {\alpha V}}+{\vec U}^R)^2 +\nonumber\\
 &&\sum_{r}
 \{ {1\over 2} ({\overline {\alpha V}}^{\, r})^2+
 \sum_{q=1}^{\infty} [(q+{\overline {\alpha V}}^{\, r} -{1\over 2} )k^r_q +
 (q-{\overline {\alpha V}}^{\, r}-{1\over 2 }){\overline k}^{\, r}_q ]\} ~.
\end{eqnarray}
Here $n^\ell_q$ and ${\overline n}^\ell_q$ are occupation numbers for the
left-moving bosons $\phi^\ell_L$,
whereas $m^\ell_q$ and ${\overline m}^\ell_q$ are those
for the right-moving bosons
$\phi^\ell_R$. These take non-negative integer values.
$k^r_q$ and ${\overline k}^{\, r}_q$ are the occupation numbers for the
right-moving fermions, and they take only two values: $0$ and $1$.
The occupation numbers are directly related to the boson and fermion number
operators. For example, $N^r_{\overline {\alpha V}} =\sum_{q=1}^{\infty}
(k^r_q -{\overline k}^{\, r}_q)$.

{}We conclude this section by summarizing the rules. To construct a consistent
orbifold model, start with an $N=4$ space-time supersymmetric four dimensional
heterotic string model with the internal momenta spanning an
even self-dual lattice
$\Gamma^{6,22}$ that possesses a ${\bf Z}_k$ symmetry
($k$ is a prime). The invariant sublattice $I$ must be such that $N_I =1$ or
$k$. Now one can introduce a set
of vectors ${V_i}$ (which includes $V_0$) that correspond to a particular
embedding of the orbifold group $Z_k$. A given embedding is acceptable if
and only if the set $\{V_i\}$ satisfies (\ref{supercurrent}), and
the set of constraints (\ref{k1}), (\ref{k2}) and (\ref{k3})
for some choices of the structure constants $k_{ij}$. Then,
a particular choice of the set $\{V_i, k_{ij}\}$ defines a consistent string
model. The complete spectrum (on- and off-shell states)
of the model is given by  the spectrum generating formula (\ref{sgf}),
which together with the left/right energy formula (\ref{LRenergy}) determines
the on-shell physical spectrum. In the next three sections we will illustrate
the rules with some examples.

%============================================================================

%\newpage
\section{Asymmetric Orbifolds and Wilson Lines}\label{secIV}
\bigskip

{}Consider an even self-dual Lorentzian lattice $\Gamma^{6,22} =
\Gamma^{2,2} \otimes \Gamma^{2,2} \otimes
\Gamma^{2,2} \otimes \Gamma^8
\otimes \Gamma^8$. Here we take $\Gamma^{2,2}$ to be the even
self-dual Lorentzian lattice spanned by the vectors
$({\overline p},p)$ such that ${\overline p},p \in
{\tilde \Gamma}^2$ ($SU(3)$ weight lattice), and
$p-{\overline p} \in \Gamma^2$; $\Gamma^2$ ($SU(3)$ root lattice).
$\Gamma^8$ is the $E_8$ root lattice.
The lattice $\Gamma^{6,22}$ has a
${\bf Z}_3$ symmetry under $120^\circ$ rotations of the right-moving momenta
while the left-moving momenta are untouched. Here we discuss asymmetric
orbifolds obtained via modding out by this discrete symmetry.

{}Consider the following set of generating vectors
\begin{eqnarray}
 &&W=(0(0~{1\over 2})^3\vert\vert 0^3 \vert 0^{8} \vert 0^8 )~,\\
 &&V_0 =(-{1\over 2} (-{1\over 2}~ 0)^3 \vert\vert
 0^3 \vert 0^{8} \vert 0^8 ) ~,\\
 &&V_1 =( 0 (-{1\over 3}~{1\over 3} )^3 \vert\vert 0^3 \vert v^I )~,\\
 &&V_2 =( 0 ( 0~0)^3 \vert\vert w~ 0^2 \vert A^I )~,\\
 &&V_3 =( 0 ( 0~0)^3 \vert\vert 0~({1\over 2}\zeta)^2 \vert {\tilde A}^I )~
\end{eqnarray}
(Here we have chosen the basis where all the right-moving fermions,
as well as the bosons corresponding
to the $\Gamma^{2,2} \otimes \Gamma^{2,2} \otimes
\Gamma^{2,2}$ sublattice, are complex; the first single vertical line
separates the three complex and sixteen real left-moving bosons, the latter
corresponding to the $\Gamma^8 \otimes \Gamma^8$ sublattice; the second
single vertical line separates the real bosons corresponding to the first and
second $\Gamma^8$ sublattices, respectively). Note that
$V_1 \cdot W ={1\over 2}$. Let us choose
$3v^I , 3A^I , 2{\tilde A}^I
\in \Gamma^{8} \otimes \Gamma^8$; $w \in {\tilde \Gamma}^2$, $w^2={2\over 3}$;
$\zeta \in \Gamma^2$, $\zeta^2 =2$. Then
$m_1 =t_1 =m_2 =3$, $m_3=2$. The matrix of the dot products $V_i \cdot V_j$
reads (for simplicity we choose ${\tilde A}^I v^I ={\tilde A}^I A^I =0$):
\begin{equation}
 V_i \cdot V_j=\left( \begin{array}{cccc}
               -1 &  -{1\over 2}& 0 & 0\\
               -{1\over 2} & (v^I)^2 & v^I A^I & 0\\
               0 & v^I A^I & (A^I)^2 +{2\over 3} & 0\\
               0 & 0 & 0 & ({\tilde A}^I)^2 +1
               \end{array}
        \right)~.
\end{equation}
The structure constants $k_{ij}$ are given by:
\begin{equation}
 k_{ij}=\left( \begin{array}{cccc}
               k_{00} & 0 & 0 & k_{30}\\
               {1\over 2} & {1\over 2}(v^I)^2 & -k_{21} +v^I A^I & 0\\
               0 & k_{21} & {1\over 2}(A^I)^2 +{1\over 3} & 0\\
               k_{30} & 0 & 0 & k_{30}+{1\over 2}({\tilde A}^I)^2 +{1\over 2}
               \end{array}
        \right)~.
\end{equation}
To satisfy the constraints (\ref{k2}) we must have:
\begin{equation}
 3v^I A^I ,({\tilde A}^I )^2 \in {\bf Z}~,~~~ 3(v^I )^2 , 3(A^I )^2 \in
 2{\bf Z}~.
\end{equation}

{}The invariant sublattice ({\em i.e.}, the sublattice of $\Gamma^{6,22}$
invariant under the twist part of $V_1$) is $I=\Gamma^2 \otimes \Gamma^2
\otimes \Gamma^2 \otimes \Gamma^8 \otimes \Gamma^8$. Note that its dual
lattice is ${\tilde I} ={\tilde \Gamma}^2 \otimes {\tilde \Gamma}^2
\otimes {\tilde \Gamma}^2 \otimes {\Gamma}^8 \otimes \Gamma^8$, and
$N_I =3 (=t_1)$. The determinant of the metric of $I$ is $M_I =3^3 =27$, and
\begin{equation}
 \xi(\alpha_1 )=M^{-{1\over 2}}_I [2\sin ({{\alpha_1 \pi}\over 3})]^3 =
 1~,~~~\alpha_1 =1,2~.
\end{equation}
Therefore, the number of fixed points in each of the twisted sectors is one.

{}Before discussing the orbifold models generated by these vectors, we note
that the model, which we will refer to as $N0$, generated by the set
$\{V_0\}$ (which contains only $V_0$)
is an $N=4$ space-time supersymmetric Narain model
\cite{Narain}. Its massless spectrum
consists of the $N=4$ supergravity multiplet (graviton, four gravitinos,
six vector bosons, four spin-${1\over 2}$ fermions, and one complex
scalar (dilaton plus axion)), and also the $N=4$ super-Yang-Mills multiplet
(gauge bosons, four spin-${1\over 2}$ fermions, and six real scalars)
transforming in the adjoint of the gauge group $E_8 \otimes E_8 \otimes
SU(3)\otimes SU(3) \otimes SU(3)$. The bosons come from the $\bf 0$ sector,
whereas their superpartners come from the $V_0$ sector.

%============================================================================
\subsection{Asymmetric Orbifolds without Wilson Lines}
\medskip

{}Next, consider the model, which we will refer to as $A1$,
generated by the set $\{V_0 ,V_1\}$,
with $v^I=({1\over 3}~{1\over 3}~{2\over 3}~0^5 \vert 0^8 )$.
This is an asymmetric orbifold model without Wilson lines. It possesses $N=1$
space-time supersymmetry. The sectors ${\overline {\alpha V}}$, $\alpha_0 =0$,
give rise to bosons, whereas their superpartners come from the sectors
${\overline {\alpha V}}$, $\alpha_0 =1$.

{}First consider the untwisted sectors ${\overline {\alpha V}}=
{{\alpha_0 V_0}}$ ($\alpha_1 =0$). In these sectors the momenta
${\vec P}
_{\overline {\alpha V}} =({\overline p}_1 ,{\overline p}_2 ,{\overline p}_3
\vert\vert
p_1 ,p_2 ,p_3 \vert p^I )$ (${\overline p}_\ell ,p_\ell
\in {\tilde \Gamma}^2$,
$p_\ell - {\overline p}_\ell
\in \Gamma^2$, $\ell=1,2,3$; $p^I \in \Gamma^8 \otimes
\Gamma^8$) span $\Gamma^{6,22}$.

{}The spectrum generating formula in the untwisted sectors reads:
\begin{eqnarray}
 &&V_0 \cdot {\cal N}_{\overline {\alpha V}}=
 {1\over 2} \sum_{r=0}^3
  N^r_{\overline {\alpha V}} =k_{00} \alpha_0 +{1\over 2} ~
 (\mbox{mod~1})~,\\
 &&V_1 \cdot {\cal N}_{\overline {\alpha V}}=
 {1\over 3} (\sum_{r=1}^3 N^r_{\overline {\alpha V}}+
 \sum_{\ell=1}^3
 J^\ell_{\overline {\alpha V}} )+v^I p^I  =0~(\mbox{mod~1})~.
\end{eqnarray}

{}Thus, the untwisted sectors
${\bf 0}$ ($\alpha_0 =0$) and $V_0$ ($\alpha_0 =1$) give rise to the following
massless states: ({\em i}) The graviton $N=1$ supermultiplet (graviton and
one gravitino); ({\em ii}) the dilaton $N=1$ supermultiplet (one
spin-${1\over 2}$ fermion and one complex scalar (dilaton plus axion));
({\em iii}) $N=1$ Yang-Mills supermultiplet
(gauge bosons and one spin-${1\over 2}$ fermion)
transforming
in the adjoint of $SU(3) \otimes E_6 \otimes E_8 \otimes (SU(3))^3$
(These states satisfy $v^I p^I \in {\bf Z}$;
the first
$SU(3)$ subgroup arises in the breaking $E_8 \supset SU(3) \otimes E_6$ due
to the shift $v^I$; the other three $SU(3)$ subgroups come from the $\Gamma
^{2,2} \otimes \Gamma^{2,2} \otimes \Gamma^{2,2}$ lattice); ({\em iv}) three
$N=1$ chiral supermultiplets transforming
in the representation $({\bf 3}, {\bf 27}, {\bf 1}
, {\bf 1}, {\bf 1}, {\bf 1})$ of the gauge group
(These states have $v^I p^I \in \pm {1\over 3} +{\bf Z}$ for those
transforming in $\bf 3$ and $\overline {\bf 3}$, respectively).
The chirality of these states depends on the choice of $k_{00}$:
For $k_{00}=1/2$ they are left-movers, whereas for
$k_{00}=0$ they are right-movers. For definiteness in the following
we choose $k_{00}=1/2$.

{}Next, consider the twisted sectors ${\overline {\alpha V}}=
{\overline {\alpha_0 V_0 +\alpha_1 V_1}}$ ($\alpha_1 =1,2$).
In these sectors the momenta
${\vec P}_{\overline {\alpha V}} =(0,0,0 \vert\vert
p_1 ,p_2 ,p_3 \vert p^I )$
($p_\ell \in {\tilde \Gamma}^2$,
$\ell=1,2,3$; $p^I \in \Gamma^8 \otimes
\Gamma^8$) span ${\tilde I}$.

{}The spectrum generating formula in the twisted sectors
reads ($(v^I)^2 =2/3$)
\begin{eqnarray}
 &&V_0 \cdot {\cal N}_{\overline {\alpha V}} =
 {1\over 2} \sum_{r=0}^3
  N^r_{\overline {\alpha V}} =\alpha_0 (k_{00} +{1\over 2}\alpha_1 )
 ~(\mbox{mod~1})~,\\
 &&\alpha_1 V_1 \cdot {\cal N}_{\overline {\alpha V}} +
 {1\over 2}{\vec P}^2_{\overline {\alpha V}}=\nonumber\\
 &&\,\,\,\,\,\,\,{\alpha_1 \over 3} (\sum_{r=1}^3 N^r_{\overline {\alpha V}}+
 \sum_{\ell=1}^3
 J^\ell_{\overline {\alpha V}})+v^I p^I+{1\over 2} \sum_{\ell=1}^3 p^2_\ell
 =-{1\over 3} \alpha^2_1 ~(\mbox{mod~1})~.
\end{eqnarray}

{}The twisted sectors give rise to one left-moving chiral $N=1$ supermultiplet
transforming in the following representations of the gauge group: ({\em i})
$({\bf 3}, {\overline {\bf 27}},
{\bf 1}, {\bf 1}, {\bf 1}, {\bf 1})$ (these have $v^I p^I \in -{1\over 3} +
{\bf Z}$); ({\em ii}) $({\bf 1}, {{\bf 27}},
{\bf 1}, {\bf x}, {\bf y}, {\bf z})$, where one of
${\bf x}, {\bf y}, {\bf z})$ is ${\bf 3}$ or ${\overline {\bf 3}}$, and the
other two are ${\bf 1}$ (these have $v^I p^I \in {1\over 3} +
{\bf Z}$); ({\em iii}) $({\overline {\bf 3}}, {{\bf 1}},
{\bf 1}, {\bf x}, {\bf y}, {\bf z})$, where two of
${\bf x}, {\bf y}, {\bf z}$ are ${\bf 3}$ or ${\overline {\bf 3}}$, and the
other one is ${\bf 1}$ (these have $v^I p^I \in {\bf Z}$).

{}The $A1$ model is the original asymmetric orbifold model of \cite{NSV}.

%===========================================================================
\subsection{Turning on Wilson Lines}\label{subsecIVB}
\medskip

{}Wilson lines are incorporated via vectors $V_i$ that contain only lattice
shifts. Thus, consider the model, which we will refer to as $N2$,
generated by the vectors $V_0$ and $V_2$, with $A^I=
({1\over 3}~{1\over 3}~{2\over 3}~0^5 \vert
(-{1\over 3})~
(-{1\over 3})~(-{2\over 3})~0^5  )$, and $w$ being a weight vector
corresponding to the $\bf 3$ irrep of $SU(3)$. This model is an $N=4$
space-time supersymmetric Narain model. The sectors with $\alpha_0=0$ give
rise to bosons, whereas their superpartners are supplied by the sectors with
$\alpha_0=1$.

{}The spectrum generating formula reads ($(A^I)^2 =4/3$):
\begin{eqnarray}
 &&V_0 \cdot {\cal N}_{\overline {\alpha V}}=
 {1\over 2} \sum_{r=0}^3
  N^r_{\overline {\alpha V}}=k_{00} \alpha_0 +{1\over 2}
 ~(\mbox{mod~1})~,\\
 &&V_2 \cdot {\cal N}_{\overline {\alpha V}}=
 wp_1 + A^I p^I=0~(\mbox{mod~1})~.
\end{eqnarray}

{}The momenta that survive this projection are given by:
\begin{equation}\label{N2}
 {\vec P}=({\overline q}+3A^I p^I w ,{\overline p}_2 ,{\overline p}_3
 \vert\vert
 q+ (3A^I p^I +\alpha_2 )w ,p_2 ,p_3 \vert p^I +\alpha_2 A^I ) ~,
\end{equation}
where ${\overline q}, q \in \Gamma^2$.

{}Thus, the ``unshifted'' sectors ($\alpha_2 =0$) give rise to
the following massless states: ({\em i}) The $N=4$ supergravity multiplet;
({\em ii}) $N=4$ vector supermultiplet transforming in the adjoint of
$SU(3) \otimes E_6 \otimes SU(3)
\otimes E_6 \otimes SU(3) \otimes (SU(3))^2$.

{}The ``shifted'' sectors ($\alpha_2 =1$ and $2$) give rise to the
$N=4$ vector supermultiplets transforming in the
representations  $({\overline{\bf 3}},{\bf 1},{\bf 3},{\bf 1},{\bf 3},{\bf 1}
,{\bf 1})$ and
$({\bf 3},{\bf 1},{\overline {\bf 3}},{\bf 1},{\overline{\bf 3}},{\bf 1},
{\bf 1})$, respectively. These states combine with those in the ``unshifted''
sectors and give rise to the $N=4$ Yang-Mills supermultiplet transforming in
the adjoint of the resulting gauge group
$E_6 \otimes E_6 \otimes E_6 \otimes SU(3) \otimes SU(3)$.

%===========================================================================
\subsection{Asymmetric Orbifolds with Wilson Lines}
\medskip

{}Now we turn to asymmetric orbifold models with Wilson lines. Consider the
model, which we will refer to as $A2$, generated by the vectors $V_0$, $V_1$
and $V_2$. This model has $N=1$ space-time supersymmetry.

{}First consider the untwisted sectors ${\overline {\alpha V}} =\alpha_0 V_0
+\alpha_2 V_2$. The spectrum generating formula reads:
\begin{eqnarray}
 &&V_0 \cdot {\cal N}_{\overline {\alpha V}}=
 {1\over 2} \sum_{r=0}^3
  N^r_{\overline {\alpha V}} =k_{00} \alpha_0 +{1\over 2} ~
 (\mbox{mod~1})~,\\
 &&V_1 \cdot {\cal N}_{\overline {\alpha V}}=
 {1\over 3} (\sum_{r=1}^3 N^r_{\overline {\alpha V}}+
 \sum_{\ell=1}^3
 J^\ell_{\overline {\alpha V}} )+v^I p^I  =-k_{21} \alpha_2 ~(\mbox{mod~1})~,\\
 &&V_2 \cdot {\cal N}_{\overline {\alpha V}}=
 wp_1 + A^I p^I=0~(\mbox{mod~1})~.
\end{eqnarray}

{}If $k_{21}=0$, then the gauge group of the $A2$ model is the same as that
of $N2$. If $k_{21}\not=0$, then the gauge group is broken down to $(E_6 )^2
\otimes (SU(3))^5$. For definiteness we will choose $k_{21}=0$. Then the
untwisted sectors contribute the following
massless states: ({\em i}) The graviton $N=1$ supermultiplet; ({\em ii}) the
dilaton $N=1$ supermultiplet; ({\em iii}) $N=1$ Yang-Mills supermultiplet
transforming in the adjoint of the gauge group
$E_6 \otimes E_6 \otimes E_6 \otimes (SU(3))^2$
(these states satisfy $v^I p^I \in {\bf Z}$).

{}Next, consider the twisted sectors ${\overline {\alpha V}}=
{\overline {\alpha_0 V_0 +\alpha_1 V_1}}$ ($\alpha_1 =1,2$).
The spectrum generating formula in the twisted sectors
reads ($(v^I)^2 =2/3$):
\begin{eqnarray}
 &&V_0 \cdot {\cal N}_{\overline {\alpha V}} =
 {1\over 2} \sum_{r=0}^3
  N^r_{\overline {\alpha V}} =\alpha_0 (k_{00} +{1\over 2}\alpha_1 )
 ~(\mbox{mod~1})~,\\
 &&\alpha_1 V_1 \cdot {\cal N}_{\overline {\alpha V}} +
 {1\over 2}{\vec P}^2_{\overline {\alpha V}}=
 {\alpha_1 \over 3} (\sum_{r=1}^3 N^r_{\overline {\alpha V}}+
 \sum_{\ell=1}^3
 J^\ell_{\overline {\alpha V}})+v^I p^I+{1\over 2} \sum_{\ell=1}^3 p^2_\ell
 =\nonumber\\
 &&\,\,\,\,\,\,\,-{1\over 3} \alpha^2_1 ~(\mbox{mod~1})~,\\
 &&V_2 \cdot {\cal N}_{\overline {\alpha V}}=
 wp_1 + A^I p^I=-{2\over 3}\alpha_1 ~(\mbox{mod~1})~.
\end{eqnarray}

{}The twisted sectors with $\alpha_2 =0$ give rise to one left-moving chiral
$N=1$ supermultiplet transforming in the following representations of
the group $SU(3)\otimes E_6 \otimes SU(3) \otimes E_6
\otimes SU(3) \otimes (SU(3))^2$:
({\em i}) $({\bf 1}, {{\bf 27}},
{\bf 1}, {\bf 1}, {\bf 1}, {\bf x}, {\bf y})$, where one of
${\bf x}, {\bf y}$ is ${\bf 3}$ or ${\overline {\bf 3}}$, and the
other one is ${\bf 1}$ (these have $A^I p^I \in {1\over 3} +
{\bf Z}$); ({\em ii}) $({\overline {\bf 3}}, {{\bf 1}},
{\bf 1}, {\bf 1}, {\overline {\bf 3}}, {\bf x}, {\bf y})$, where one of
${\bf x}, {\bf y}$ is ${\bf 3}$ or ${\overline {\bf 3}}$, and the
other one is ${\bf 1}$ (these have $A^I p^I \in {\bf Z}$).

{}The twisted sectors with $\alpha_2 =2$ give rise to one left-moving chiral
$N=1$ supermultiplet transforming in the following representations of
the group $SU(3)\otimes E_6 \otimes SU(3) \otimes E_6
\otimes SU(3) \otimes (SU(3))^2$:
({\em i}) $({\bf 1}, {{\bf 1}},
{\bf 1}, {\bf 27}, {\bf 1}, {\bf x}, {\bf y})$, where one of
${\bf x}, {\bf y}$ is ${\bf 3}$ or ${\overline {\bf 3}}$, and the
other one is ${\bf 1}$ (these have $A^I p^I \in -{1\over 3} +
{\bf Z}$); ({\em ii}) $({{\bf 1}}, {{\bf 1}},
{\overline {\bf 3}}, {\bf 1},
{{\bf 3}}, {\bf x}, {\bf y})$, where one of
${\bf x}, {\bf y}$ is ${\bf 3}$ or ${\overline {\bf 3}}$, and the
other one is ${\bf 1}$ (these have $A^I p^I \in {\bf Z}$).

{}The twisted sectors with $\alpha_2 =1$ give rise to one left-moving chiral
$N=1$ supermultiplet transforming in the following representations of
the group $SU(3)\otimes E_6 \otimes SU(3) \otimes E_6
\otimes SU(3) \otimes (SU(3))^2$:
$({\bf 3}, {{\bf 1}},
{\bf 3}, {\bf 1}, {\bf 1}, {\bf x}, {\bf y})$, where one of
${\bf x}, {\bf y}$ is ${\bf 3}$ or ${\overline {\bf 3}}$, and the
other one is ${\bf 1}$ (these have $A^I p^I \in
{\bf Z}$).

{}Thus, the states from the twisted sectors combine into the following
representations of the resulting gauge group $E_6 \otimes E_6 \otimes E_6
\otimes SU(3) \otimes SU(3)$: There is one left-moving
chiral $N=1$ supermultiplet in the representations
$({\bf 27}, {\bf 1}, {\bf 1}, {\bf x}, {\bf y})$,
$({\bf 1}, {\bf 27}, {\bf 1}, {\bf x}, {\bf y})$,
$({\bf 1}, {\bf 1}, {\bf 27}, {\bf x}, {\bf y})$,
where one of
${\bf x}, {\bf y}$ is ${\bf 3}$ or ${\overline {\bf 3}}$, and the
other one is ${\bf 1}$.

{}Finally, we briefly discuss the model, which we refer to as $A3$, generated
by the vectors $V_0$, $V_1$, $V_2$ and $V_3$, with ${\tilde A}^I =
(0^7 ~1 \vert 0^7 ~1)$. Note that if $k_{30}=1/2$, the supersymmetry is broken
to $N=0$.  For definiteness we will choose $k_{30}=0$. Then the model possesses
$N=1$ space-time supersymmetry. The sectors ${\overline {\alpha V}}$ with
$\alpha_3 =1$ contribute massive string states only. However,
some of the states from the other sectors are projected out due to the presence
of the Wilson line generated by $V_3$. Thus, the gauge symmetry is broken down
to $E_6 \otimes (SO(10) \otimes U(1))^2 \otimes (SU(2)\otimes U(1))^2$.
In the twisted sectors we have chiral $N=1$ supermultiplets
in the following
representations of $E_6 \otimes SO(10) \otimes SO(10) \otimes SU(2)
\otimes SU(2)$
(Here we drop the $U(1)$ charges for the sake of
simplicity; they are straightforward to work out, however): Four fields
(with different $U(1)$ charges) in
$({\bf 27}, {\bf 1}, {\bf 1}, {\bf 1}, {\bf 1})$; one field in each of
$({\bf 1}, {\bf 16}, {\bf 1}, {\bf 2}, {\bf 1})$,
$({\bf 1}, {\bf 16}, {\bf 1}, {\bf 1}, {\bf 2})$,
$({\bf 1}, {\bf 1}, {\bf 16}, {\bf 2}, {\bf 1})$,
$({\bf 1}, {\bf 1}, {\bf 16}, {\bf 1}, {\bf 2})$. We also have Higgs $N=1$
supermultiplets in the following representations: Four fields (with
different $U(1)$ charges) in each of
$({\bf 1}, {\bf 10}, {\bf 1}, {\bf 1}, {\bf 1})$,
$({\bf 1}, {\bf 1}, {\bf 10}, {\bf 1}, {\bf 1})$. There are also four
singlets in the twisted sectors.

{}The examples we have considered indicate that incorporating Wilson lines
into asymmetric orbifolds may be useful for controlling the gauge symmetry and
the number of chiral generations in a given model. As we discuss in the
next section, asymmetric orbifolds are also very handy in constructing models
with reduced rank.

%============================================================================

%\newpage
\section{Rank Reduction}\label{secV}
\bigskip

{}{}We start with an
even self-dual Lorentzian lattice $\Gamma^{6,22} =
\Gamma^{2,2} \otimes \Gamma^{2,2} \otimes
\Gamma^{2,2} \otimes \Gamma^8
\otimes \Gamma^8$. Here we take $\Gamma^{2,2}$ to be the even
self-dual Lorentzian lattice spanned by the vectors
$({\overline p},p)$ such that ${\overline p},p \in
{\tilde \Gamma}^2$ ($SU(3)$ weight lattice), and
$p-{\overline p} \in \Gamma^2$ ($SU(3)$ root lattice).
$\Gamma^8$ is the $E_8$ root lattice.

{}Consider the following set of generating vectors (Here
we choose $\zeta\in\Gamma^2$, $\zeta^2 =2$).
\begin{eqnarray}
 &&W=(0 (0~{1\over 2})^2 (0~0) \vert\vert ({1\over 2})^2 ~0\vert
 ({1\over 2})^4 \vert 0^4 )~,\\
 &&V_0 =(-{1\over 2} (-{1\over 2}~ 0)^3\vert 0^3\vert 0^{4}\vert 0^4 )~,\\
 &&V_1 =(0(-{1\over 2}~ {1\over 2})^2 (0~0)\vert
 ({1\over 2} )^2 ({1\over 2}\zeta)\vert ({1\over 2})^4 \vert 0^4)~.
\end{eqnarray}

{}Here all the bosons are complexified. The complexification for the sixteen
$E_8 \otimes E_8$ real bosons $\varphi^I_1$ and $\varphi^I_2$, $I=1,...,8$
is chosen as $\phi^7={1\over 2}(\varphi^1_1-\varphi^1_2 +i
(\varphi^2_1-\varphi^2_2)),...,
\phi^{10}={1\over 2}(\varphi^7_1-\varphi^7_2 +i
(\varphi^8_1-\varphi^8_2))$;
$\phi^{11}={1\over 2}(\varphi^1_1+\varphi^1_2 +i
(\varphi^2_1+\varphi^2_2)),...,
\phi^{14}={1\over 2}(\varphi^7_1+\varphi^7_2 +i
(\varphi^8_1+\varphi^8_2))$. The
second single vertical line in
$V_i$  separates $\phi^\ell$, $\ell=7,...,10$, from $\phi^\ell$,
$\ell=11,...,14$.

{}The twist $V_1$ has the following action on $\phi^\ell$, $\ell=7,...,14$:
$\phi^\ell \rightarrow -\phi^\ell$, $\ell=7,...,10$, and
$\phi^\ell \rightarrow \phi^\ell$, $\ell=11,...,14$. This corresponds
to modding out by the permutational symmetry $\phi^I_1
\leftrightarrow \phi^I_2$, that is, the outer automorphism of the $\Gamma^8
\otimes \Gamma^8$ lattice.

{}The matrix of the dot products $V_i \cdot V_j$ and
structure constants $k_{ij}$ for this model are given by
\begin{equation}
 V_i \cdot V_j =\left( \begin{array}{cc}
               -1& -{1\over 2} \\
               -{1\over 2} & -1
               \end{array}
        \right)~,~~~
 k_{ij}=\left( \begin{array}{cc}
               k_{00} & {k_{10}+{1\over 2}} \\
               k_{10} & {k_{10}+{1\over 2}}
               \end{array}
        \right)~.
\end{equation}

{}The invariant sublattice is $I =\Gamma^{2,2} \otimes
\Delta^8$, where ${\Delta}^8 \equiv \{\sqrt{2} p^I \vert p^I\in
{\Gamma}^8 \}$. The dual lattice is ${\tilde I}= \Gamma^{2,2} \otimes
{\tilde \Delta}^8$, where
${\tilde \Delta}^8 \equiv \{{1\over \sqrt{2}} p^I \vert p^I\in
{\Gamma}^8 \}$ is the lattice dual to $\Delta^8$. Note that $N_I=2(=t_1)$.
The determinant of the metric of $I$ is $M_I=2^8$, and in the twisted sectors
($\alpha_1 =1$) we have
\begin{equation}
 \xi(\alpha_1 )=M^{-{1\over 2}}_I [2\sin ({{\alpha_1 \pi}\over 2})]^8 =
 16~.
\end{equation}

{}The model generated by the set $\{V_0 ,V_1 \}$ has $N=2$ space-time
supersymmetry. The bosons come from the sectors with $\alpha_0 =0$, whereas the
fermions arise in the sectors with $\alpha_0=1$.

{}First consider the
untwisted sectors ${\overline {\alpha V}}=\alpha_0 V_0$ ($\alpha_1 =0$).
In these sectors the momenta ${\vec P}
_{\overline {\alpha V}} =({\overline p}_1 ,{\overline p}_2 ,{\overline p}_3
\vert\vert
p_1 ,p_2 ,p_3 \vert q )$ (${\overline p}_\ell ,p_\ell
\in {\tilde \Gamma}^2$,
$p_\ell - {\overline p}_\ell
\in \Gamma^2$, $\ell=1,2,3$; $q \in \Gamma^8 \otimes
\Gamma^8$) span $\Gamma^{6,22}$.

{}The spectrum generating formula in the untwisted sectors reads
\begin{eqnarray}
 &&V_0 \cdot {\cal N}_{\overline {\alpha V}}
 ={1\over 2} \sum_{r=0}^3 N^r_{\overline {\alpha V}}
  =k_{00} \alpha_0 +{1\over 2}
 ~(\mbox{mod}~1)~,\\
 &&V_1 \cdot {\cal N}_{\overline {\alpha V}}=
 {1\over 2}(N^1_{\overline {\alpha V}} +N^2_{\overline {\alpha V}}
 +J^1_{\overline {\alpha V}} +J^2_{\overline {\alpha V}}
 -J^4_{\overline {\alpha V}} -J^5_{\overline {\alpha V}}-\nonumber\\
 &&~~~~~~~-\sum_{\ell=7}^{10} J^\ell_{\overline {\alpha V}} +
 \zeta p_3 )=(k_{10}+{1\over 2} )\alpha_0 ~(\mbox{mod}~1)~.
\end{eqnarray}

{}The untwisted sectors
${\bf 0}$ and $V_0$ give rise to the following
massless states:
({\em i}) The $N=2$ supergravity multiplet; ({\em ii}) $N=2$ Yang-Mills
supermultiplet transforming in the adjoint of the gauge group $E_8
\otimes SU(2) \otimes SU(2) \otimes (SU(2) \otimes U(1))$ gauge group;
({\em iii}) two $N=2$ scalar supermultiplets in each of the
representations $({\bf 248}, {\bf 1}, {\bf 1}, {\bf 1})(0)$, $({\bf 1},
{\bf 5}, {\bf 1}, {\bf 1})(0)$, $({\bf 1},{\bf 1},{\bf 5},{\bf 1})(0)$,
$({\bf 1},{\bf 1},{\bf 1},{\bf 2})(3)$ and
$({\bf 1},{\bf 1},{\bf 1},{\bf 2})(-3)$ of $E_8
\otimes SU(2) \otimes SU(2) \otimes (SU(2) \otimes U(1))$ (The $U(1)$ charge
is given in regular font in the parentheses).

{}In the twisted sectors ${\overline {\alpha V}}={\overline
{\alpha_0 V_0 +V_1}}$ the momenta
${\vec P}_{\overline {\alpha V}}=(0,0,{\overline p}_3 \vert\vert 0,0, p_3 \vert
0^4 \vert Q )$ (${\overline p}_3 ,p_3 \in {\tilde \Gamma}^2$,
$p_3-{\overline p}_3 \in \Gamma^2$; $Q \in {\tilde \Delta}^8$) span
${\tilde I}$.

{}The spectrum generating formula in the twisted sectors reads ($\alpha_1 =1$):
\begin{eqnarray}
&&V_0 \cdot {\cal N}_{\overline {\alpha V}}
 ={1\over 2} \sum_{r=0}^3 N^r_{\overline {\alpha V}}
  =k_{00} \alpha_0 +k_{10}+{1\over 2}
 ~(\mbox{mod}~1)~,\\
 &&V_1 \cdot {\cal N}_{\overline {\alpha V}}
 +{1\over 2}{\vec P}^2_{\overline {\alpha V}}=
 {1\over 2}(N^1_{\overline {\alpha V}} +N^2_{\overline {\alpha V}}
 +J^1_{\overline {\alpha V}} +J^2_{\overline {\alpha V}}
 -J^4_{\overline {\alpha V}} -J^5_{\overline {\alpha V}}-\nonumber\\
 &&~~~~~~~-\sum_{\ell=7}^{10} J^\ell_{\overline {\alpha V}} +
 \zeta p_3 +Q^2)=(k_{10}+{1\over 2} )(\alpha_0+1) ~(\mbox{mod}~1)~.
\end{eqnarray}

{}Thus, the twisted sectors give rise to the following massless states:
Four $N=2$ scalar supermultiplets
transforming in the representation $({\bf 1}, {\bf 2}, {\bf 2}, {\bf 2}) (0)$
of the gauge group.

{}Note the rank reduction of the gauge group from twenty-two to twelve. This
indicates that the gauge group is realized via a higher level Kac-Moody
algebra. Also note appearance of massless states in ${\bf 5}$ irrep of
$SU(2)$. This is too a sign of a higher level Kac-Moody algebra realization.
A careful analysis of the underlying conformal field theory unambiguously
determines the levels of each subgroup: $E_8$ is realized at level two
(It arises in the breaking $(E_8)_1 \otimes (E_8)_1 \supset (E_8)_2 \times
(\mbox{Ising Model})$;
the central charge of $(E_8)_2$ is $c=15/2$,
whereas an Ising model has $c=1/2$);
the first two $SU(2)$ subgroups are realized at
level four (as a result of a special breaking $SU(3)_1 \supset SU(2)_4$; note
that under this breaking ${\bf 8}={\bf 3}+{\bf 5}$, and the central charge
of both $SU(3)_1$ and $SU(2)_4$ is $c=2$); the last $SU(2)$ is realized at
level one (It arises in a regular breaking $SU(3)_1 \supset SU(2)_1 \otimes
U(1)$).

%============================================================================

%\newpage
\section{Other Examples}\label{secVI}
\bigskip

{}For completeness, in this section we briefly discuss
symmetric orbifolds with and
without Wilson lines. We construct these examples
(familiar from the previous developments \cite{DHVW,INQ}) using the rules
given in section \ref{secIII}. This is to further clarify the rules and notation.

\subsection{Symmetric Orbifolds without Wilson Lines}
\medskip

{}We start from the lattice $\Gamma^{6,22}=\Gamma^{2,2} \otimes \Gamma^{2,2}
\otimes \Gamma^{2,2} \otimes \Gamma^{8} \otimes \Gamma^8$
considered in the previous section.
Consider the model generated by the following set of vectors
\begin{eqnarray}
 &&W   =( 0 (0~{1\over 2})^3 \vert\vert ({1\over 2})^3 \vert 0^{8} \vert
 0^8 )~,\\
 &&V_0 =(-{1\over 2} (-{1\over 2}~ 0)^3
 \vert\vert 0^3 \vert 0^{8} \vert 0^8)~,\\
 &&V_1 =( 0 ( -{1\over 3}~{1\over 3})^3 \vert\vert ({1\over 3})^3 \vert
 v^I)~,
\end{eqnarray}
where the order of the shift $v^I$ is three, {\em i.e.}, $3v^I \in
\Gamma^{8}\otimes \Gamma^8$. The matrix
of the dot products $V_i \cdot V_j$ and structure
constants $k_{ij}$ are given by
\begin{equation}
 V_i \cdot V_j =\left( \begin{array}{cc}
               -1 & -{1\over 2} \\
               -{1\over 2} & (v^I)^2 -{1\over 3}
               \end{array}
        \right)~,~~~
 k_{ij}=\left( \begin{array}{cc}
               k_{00} & 0 \\
               {1\over 2} & {1\over 2}(v^I)^2 +{1\over 3}
               \end{array}
        \right)~.
\end{equation}
Due to the constraint (\ref{k2}) we have $m_1 k_{11} \in {\bf Z}$
($m_1 =t_1=3$),
and the shift $v^I$ must satisfy the condition
\begin{equation}
 3(v^I )^2 \in 2{\bf Z}~.
\end{equation}

{}The sublattice $I \subset \Gamma^{6,22}$ invariant under the twist part of
$V_1$ is given by $I = \Gamma^{8} \otimes \Gamma^8$, and $N_I=M_I=1$
(since it is an even self-dual lattice). Therefore, the multiplicity of states
$\xi(\alpha)$ in the twisted sectors ($\alpha=1,2$) is given by
\begin{equation}
 \xi(\alpha_1 )=M^{-{1\over 2}}_I [2\sin ({{\alpha_1 \pi}\over 3})]^6 =
 27~.
\end{equation}

{}If we take
$v^I =({1\over 3}~{1\over 3}~{2\over 3}~0^5 \vert 0^8 )$,
then we obtain the original symmetric orbifold model of \cite{DHVW}. The gauge
group of this model is $E_6 \otimes SU(3) \otimes E_8 \otimes (U(1))^6$ (The
six $U(1)$'s survive because before orbifolding the $\Gamma^{2,2}$ sublattices
were at the special radius of enhanced gauge symmetry; after orbifolding the
original $SU(3)$ gauge group undergoes a regular breaking $SU(3)\supset U(1)
\otimes U(1)$). The twisted sectors give rise to $\xi(\alpha_1 ) =27$ chiral
matter fields (Which is the number of fixed points of the original
$Z$-orbifold).

\subsection{Symmetric Orbifolds with Wilson Lines}
\medskip

{}Now let us start from the lattice $\Gamma^{6,22}$ of the $N2$
Narain model discussed in subsection \ref{subsecIVB}. Consider the model generated by the
same vectors as in the previous subsection (The structure constants are then
the same as for the previous model).

{}The invariant sublattice is given by $I=\{ p^I \vert p^I A^I \in{\bf Z},
p^I \in \Gamma^8 \otimes \Gamma^8
\}$ (Here $A^I=({1\over 3}~{1\over 3}~{2\over 3}~0^5
\vert (-{1\over 3})~(-{1\over 3})~(-{2\over 3})~0^5)$ is the Wilson line).
The dual lattice is ${\tilde I}=\{p^I +A^I \vert
p^I \in \Gamma^8 \otimes
\Gamma^8 \}$, and $N_I=3(=t_1)$. The determinant of the
metric of $I$ is $M_I =9$, and
\begin{equation}
 \xi(\alpha_1 )=M^{-{1\over 2}}_I [2\sin ({{\alpha_1 \pi}\over 3})]^6 =
 9~, ~~~\alpha_1=1,2~.
\end{equation}

{}This model is one of the symmetric orbifold models with one Wilson line
\cite{INQ} at the special radius. The model has $N=1$ space-time
supersymmetry, and the gauge group is $E_6 \otimes E_6 \otimes (SU(6) \otimes
U(1)) \otimes (U(1))^4$ (Recall that the $N2$ model has the gauge group
$(E_6)^3 \otimes SU(3)^2$; after orbifolding one of the $E_6$ subgroups
undergoes a regular breaking $E_6 \supset SU(6) \otimes U(1)$, and each of the
$SU(3)$'s breaks to $U(1)^2$). The twisted sectors contribute chiral matter
fields with the multiplicity $\xi(\alpha_1 )=9$.

%============================================================================

%\newpage
\section{Discussion and Remarks}\label{secVII}
\bigskip

{}We have seen that the rules for asymmetric orbifolds are almost
as easy to use as those for the free fermionic string models. The rules
for asymmetric orbifold model-building are summarized at the end of
section \ref{secIII}. These rules are for models with only one twist. Can the
rules given in this paper be generalized to models with multi-twists?
In cases where the twists do not overlap, this generalization
is straightforward.
In more general situations, the twist (plus shift) operators typically
do not commute with each other. Can the rules given in this paper be
further generalized to include such non-Abelian orbifolds?

{}Recall the rules for the free fermionic string models \cite{KLT}.
It is well-known that many of the free fermionic string models, in
particular, the ones with reduced rank gauge groups, can be constructed
in the bosonic language as non-Abelian orbifolds.
However, it is also known that some of the Abelian ({\em e.g.},
${\bf Z}_2$) orbifold models
easily obtainable in the bosonic language cannot be constructed with the
free fermionic string model rules.

{}Consider the example of a single boson. By the free fermionic
string construction rules, we must start with a complex fermion, i.e.,
a boson compactified at radius one. Using the free fermionic
string rules, it is easy to change the radius to any rational value.
It is also easy to construct its ${\bf Z}_2$ orbifold at radius one,
using the real fermion basis. However, the rules \cite{KLST} do not
allow us to construct the ${\bf Z}_2$ orbifold version
at any other radius. This is because this particular model involves
a subgroup of the space group, which is non-Abelian. Fortunately, it
turns out that the free fermionic string construction
rules can be generalized to include such non-Abelian
orbifolds. Such a generalization can also be applied to the rules
given in this paper as well. This generalization will be discussed
elsewhere.

\section*{Acknowledgments}

{}This work was supported in part by the National Science Foundation.

%============================================================================
\appendix
\section{Fermion and Boson Characters}\label{appA}
\bigskip

\subsection{Free Fermions}
\medskip

{}Consider a single free left-moving complex fermion with the monodromy
\begin{equation}\label{monofer}
 \psi_v (z e^{2\pi i} )=e^{-2\pi i v}\psi_v (z)~,~~~-{1\over{2}}\leq v<
 {1\over{2}}~.
\end{equation}
The field $\psi_v (z)$ has the following mode expansion
\begin{equation}
 \psi_v (z) =\sum_{n=1}^{\infty} \{ b_{n+v-1/2} z^{-(n+v)} +
 d^{\dagger}_{n-v-1/2} z^{n-v-1} \}~.
\end{equation}
Here $b^{\dagger}_r$ and $d^{\dagger}_s$ are creation operators, and $b_r$ and
$d_s$ are annihilation operators. The quantization conditions read
\begin{equation}
 \{ b^{\dagger}_r ,b_{r^\prime} \}=\delta_{r r^\prime} ,~~~
 \{ d^{\dagger}_s ,d_{s^\prime} \}=\delta_{s s^\prime} ,~~~
 \mbox{others vanish}.
\end{equation}
The Hamiltonian $H_v$ and fermion number operator $N_v$ are
given by
\begin{eqnarray}
 &&H_v ={v^2 \over{2}}-{1\over{24}} + \nonumber\\
 &&\,\,\,\,\,\,\,\sum_{n=1}^{\infty} \{ (n+v-{1\over 2})
   b^{\dagger}_{n+v-1/2} b_{n+v-1/2}
   +(n-v-{1\over 2})d^{\dagger}_{n-v-1/2} d_{n-v-1/2} \}
   ~,\\
 &&N_v=\sum_{n=1}^{\infty} \{ b^{\dagger}_{n+v-1/2} b_{n+v-1/2}
   -d^{\dagger}_{n-v-1/2} d_{n-v-1/2} \} ~.
\end{eqnarray}
Note that the vacuum energy is ${v^2 \over{2}}-{1\over{24}}$.
Also note that for a Ramond fermion ($v=-1/2$) the vacuum is
degenerate: There are two ground states $\vert 0\rangle$ and $b^{\dagger}_0
\vert 0\rangle$.

{}The operator $-N_v$ is the generator of $U(1)$ rotations.
The corresponding characters read
\begin{eqnarray}\label{fermionZ}
 Z^v_u =&&\mbox{Tr}(q^{H_v} g^{-1} (u))= \mbox{Tr}(q^{H_v} \exp
 (-2\pi i u N_v ))= \nonumber\\
 &&q^{{v^2 \over{2}}-{1\over{24}}} \prod_{n=1}^{\infty} (1+q^{n+v-1/2}
 e^{-2\pi i u} ) (1+q^{n-v-1/2} e^{2\pi i u} ) ~.
\end{eqnarray}

{}Under the generators of modular transformations ($q=\exp(2\pi i \tau$))
\begin{equation}
 S:\tau\rightarrow -1/\tau ~,~~~T:\tau\rightarrow \tau+1~,
\end{equation}
the characters (\ref{fermionZ}) transform as
\begin{eqnarray}
 &&Z^v_u \stackrel{S}{\rightarrow} e^{2\pi ivu} Z^u_{-v} ~,\\
 &&Z^v_u \stackrel{T}{\rightarrow} e^{2\pi i({v^2 \over{2}}-{1\over{24}})}
   Z^v_{u-v-1/2} ~.
\end{eqnarray}

{}In the cases where $v=-1/2$ (Ramond fermion) or
$v=0$ (Neveu-Schwarz fermion) in (\ref{monofer}) the complex fermion $\psi_v
(z)$ can be represented in terms of a linear combination of two real
fermions. The corresponding characters for real fermions then are square roots
of the characters $Z^v_u$ for the complex fermions ($v$ and $u$ being
$-1/2$ or $0$). A more detailed discussion of the real fermion characters
is given in \cite{KLST}.

%==============================================================================
\subsection{Twisted Bosons}
\medskip

{}Consider a single free left-moving complex boson with the monodromy
\begin{equation}\label{monobos}
 \partial \phi_v (z e^{2\pi i} )=e^{-2\pi i v}\partial \phi_v (z)~,
 ~~~0\leq v<1~.
\end{equation}
The field $\partial \phi_v (z)$ has the following mode expansion
\begin{eqnarray}
 i\partial \phi_v (z) =&&\delta_{v,0} p z^{-1} + (1-\delta_{v,0} ) \sqrt{v}
 \, b_v z^{-v-1} + \nonumber\\
 &&\sum_{n=1}^{\infty} \{ {\sqrt{n+v}}\,
 b_{n+v} z^{-n-v-1} +{\sqrt{n-v}}\, d^{\dagger}_{n-v} z^{n-v-1} \}~.
\end{eqnarray}
Here $b^{\dagger}_r$ and $d^{\dagger}_s$ are creation operators, and $b_r$ and
$d_s$ are annihilation operators. The quantization conditions read
\begin{equation}
 [ b_r ,b^{\dagger}_{r^\prime} ]=\delta_{r r^\prime} ,~~~
 [ d_s ,d^{\dagger}_{s^\prime} ]=\delta_{s s^\prime} ,~~~
 [x^{\dagger} ,p]=[x,p^{\dagger} ]=i,~~~
 \mbox{others vanish}.
\end{equation}
The Hamiltonian $H_v$ and angular momentum operator $J_v$ are given by
\begin{eqnarray}
 H_v& = &\delta_{v,0} pp^{\dagger} + (1-\delta_{v,0})vb^{\dagger}_{v} b_{v} +
 \sum_{n=1}^{\infty} \{ (n+v)
   b^{\dagger}_{n+v} b_{n+v}
   +(n-v)d^{\dagger}_{n-v} d_{n-v} \} +\nonumber\\
 &&{v(1-v) \over{2}}-{1\over{12}} ~,\\
 J_v &=&\delta_{v,0}i(xp^{\dagger}-x^{\dagger} p)-
 (1-\delta_{v,0})b^{\dagger}_{v} b_{v} -
   \sum_{n=1}^{\infty} \{ b^{\dagger}_{n+v} b_{n+v}
   -d^{\dagger}_{n-v} d_{n-v} \} ~.
\end{eqnarray}
Note that the vacuum energy is ${v(1-v) \over{2}}-{1\over{24}}$.

{}The operator $J_v$
is the generator of $U(1)$ rotations.
The corresponding characters read ($v+u\not=0$):
\begin{eqnarray}\label{bosonX}
 X^v_u =&&\mbox{Tr}(q^{H_v} g^{-1} (u))= \mbox{Tr}(q^{H_v} \exp
 (2\pi i u J_v ))= \nonumber\\
 &&q^{{v(1-v) \over{2}}-{1\over{12}}}
 (1-(1-\delta_{v,0} ) q^v e^{-2\pi i u} )^{-1} \times\nonumber\\
 &&\times\prod_{n=1}^{\infty} (1-q^{n+v}
 e^{-2\pi i u} )^{-1} (1-q^{n-v} e^{2\pi i u} )^{-1} ~.
\end{eqnarray}

{}Under the generators of modular transformations
the characters (\ref{bosonX}) transform as
\begin{eqnarray}
 X^v_u &\stackrel{S}{\rightarrow}&
 ( 2\sin(\pi u) \delta_{v,0} +[2\sin(\pi v)]^{-1}
  \delta_{u,0} + (1-\delta_{vu,0})
 e^{-2\pi i(v-1/2)(u-1/2)} ) \times \nonumber\\
 &&\times X^u_{-v} ~,\\
 X^v_u &\stackrel{T}{\rightarrow}&
 e^{2\pi i({{v(1-v)}\over 2} -{1\over{12}})} X^v_{u-v} ~.
\end{eqnarray}

{}In the cases where $v=-1/2$ or $v=0$ in (\ref{monobos}),
the complex boson  $\phi_v
(z)$ can be represented in terms of a linear combination of two real
bosons. The corresponding characters for real bosons then are square roots
of the characters $X^v_u$ for the complex bosons ($v$ and $u$ being
$-1/2$ or $0$). The twisted boson characters with a different overall
normalization were discussed in Ref. \cite{RELATE}.

%==============================================================================
\subsection{Chiral Lattices}
\medskip

{}Consider $d$ free left-moving real bosons with the monodromy
\begin{equation}
 \phi^I_v (z e^{2\pi i} )=\phi^I_v (z) + v^I ~,
\end{equation}
where $I=1,2,...,d$, and $v^I$ is the $I^{\rm{\scriptstyle{th}}}$ component of the shift
vector $v$.
The field $\phi^I_v (z)$ has the following mode expansion:
\begin{equation}
 i\phi^I_v (z) =ix^I+(p^I +v^I)\ln(z) -\sum_{n\not=0} {1\over{\sqrt{n}}}
 a^I_{n} z^{-n} ~.
\end{equation}
Here $a^I_n,~n>0$, are annihilation operators, and $a^I_n,~n<0$, are creation
operators. In the following
the eigenvalues of the momentum operator $p^I$ are taken to
span an even lattice $\Gamma^d$.
The quantization conditions read
\begin{equation}
 [ a^I_n ,a^J_{n^\prime} ]=\delta^{IJ} \delta_{n n^\prime} ,
 ~~~[x^I ,p^J]=i\delta^{IJ},~~~
 \mbox{others vanish}.
\end{equation}
The Hamiltonian operator is given by
\begin{equation}
 H_v ={(p+v)^2 \over 2} + \sum_{n=1}^{\infty} n
 a^I_{-n} a^I_{n} -{d\over{24}}~.
\end{equation}

{}The momentum operator $p$ is the generator of translations. Thus, the action
of the operator
\begin{equation}
 g(u) \equiv\exp(2\pi i pu)~,~~~pu\equiv p^I u^I ~,
\end{equation}
on the field $\phi^I_v (z)$ is given by
\begin{equation}
 g(u) \phi^I_v (z) g^{-1} (u) =\phi^I_v (z) + u^I ~.
\end{equation}
The corresponding characters read
\begin{eqnarray}\label{Y_characters}
 Y^v_u =&&\mbox{Tr} (q^{H_v}
 g^{-1} (u))= \mbox{Tr} (q^{H_v} \exp(-2\pi i pu)) =\nonumber\\
 &&{1\over{\eta^d (q)}} \sum_{p\in\Gamma^d} q^{{1\over2}(p+v)^2}
 \exp(-2\pi i pu)~.
\end{eqnarray}

{}Let $w_a \in{\tilde\Gamma}^d,~a=1,...,M-1$, be a
set of vectors such that $\Gamma^d_0 \oplus \Gamma^d_1 \oplus ...
\oplus \Gamma^d_{M-1} ={\tilde \Gamma}^d$, where $w_0$ is the null vector
($w^I_0 \equiv 0$), and $\Gamma^d_a \equiv \{w_a +p \vert p\in\Gamma^d\}$,
$a=0,1,...,M-1$
(Thus, $w_a \notin
\Gamma^d$ for $a\not=0$; also note  that $M=\det(g_{ij})$).
Consider the set of characters $Y^{v+w_a }_u$:
\begin{eqnarray}\label{chiral_characters}
 &&Y^{v+w_a}_u \stackrel{T}{\rightarrow} \exp(2\pi i({1\over2} (w_a +v)^2
 -{d\over{24}})) Y^{v+w_a}_{u-v} ~,\\
 &&Y^{v+w_a}_u \stackrel{S}{\rightarrow} \sum_{b=0}^{M-1}
 S_{ab} (v,u)Y^{u+w_b}_{-v} ~,
\end{eqnarray}
where
\begin{equation}
 S_{ab} (v,u)=
%[\det(g_{ij} )]^{-1/2}
 M^{-{1\over 2}} \exp(2\pi i (w_a +v )(w_b +u))~.
\end{equation}

{}Let $N$ be the
smallest positive integer such that $\forall a~Nw^2_a \in 2{\bf Z}$.
If $N=1$ (in which case $\Gamma^d$ is an even self-dual lattice
with $M=\det(g_{ij})=1$),
we will use the characters ${Y}^v_u$ defined in (\ref{Y_characters})
whose modular transformations are particularly simple for $N=1$:
\begin{eqnarray}\label{self-dual}
 &&{Y}^v_u \stackrel{T}{\rightarrow} \exp(2\pi i({1\over2} v^2
 -{d\over{24}})) {Y}^{v}_{u-v} ~,\\
 &&{Y}^{v}_u \stackrel{S}{\rightarrow} \exp(2\pi ivu) {Y}^u_{-v} ~.
\end{eqnarray}

{}If $N>1$, the set of characters $Y^{v+w_a}_u$ is such that the
$T$-transformation is diagonal (with respect to $a$),
whereas the $S$-transformation is not.
There exists a basis such that
both $S$- and $T$-transformations act as permutations. In particular,
consider the case where $N$ is a prime.
In the discussion of asymmetric orbifolds we will use the set of characters
\begin{eqnarray}\label{CHIRAL_CHARACTERS}
 &&Y^{0, v}_{\sigma , u} \equiv Y^v_u ~,\\
 &&Y^{\lambda , v}_{\sigma , u} \equiv \sum_{a=0}^{M-1}
 \exp(-2\pi i \lambda (uw_a +{1\over2}\sigma w^2_a )) Y^{v+\lambda w_a}_u ~,
 ~~~\lambda\not=0~,
\end{eqnarray}
where $\lambda$ and $\sigma$ are integers taking values between $0$ and $N-1$,
such that $\lambda+\sigma\not=0$.
The modular transformation properties of $Y^{\lambda , v}_{\sigma , u}$ read
\begin{eqnarray}
 &&Y^{\lambda , v}_{\sigma , u}
 \stackrel{T}{\rightarrow} \exp(2\pi i({1\over 2}v^2 -{d\over{24}}))
 Y^{\lambda , v}_{\sigma-\lambda , u-v}~,\\
 &&Y^{\lambda , v}_{\sigma , u}
 \stackrel{S}{\rightarrow} \{M^{-{1\over 2}}
%[\det(g_{ij} )]^{-1/2}
\delta_{\lambda ,0} +
%[\det(g_{ij} )]^{1/2}
 M^{1\over 2} \delta_{\sigma ,0} + (1-\delta_{\lambda\sigma,0})
 \exp(2\pi i \chi(\lambda, \sigma))\}
 \exp(2\pi i vu) \times\nonumber\\
 &&\,\,\,\,\,\,\,\times Y^{\sigma , u}_{-\lambda , -v}~,
\end{eqnarray}
where $Y^{\lambda , v}_{\sigma , u} \equiv Y^{\lambda+N , v}_
{\sigma , u} \equiv Y^{\lambda , v}_{\sigma+N  , u}$, and
\begin{equation}
 \exp(2\pi i \chi(\lambda, \sigma) )
 \equiv
%[\det(g_{ij} )]^{-1/2}
 M^{-{1\over 2}}\sum_{a=0}^{M-1}
 \exp(-2\pi i {1\over 2} \lambda \sigma w^2_a ) ~,~~~\lambda\sigma\not=0 ~.
\end{equation}
Note that $\chi(\lambda, \sigma)$ are real numbers, and $\chi(\lambda, \lambda)
\equiv -d/8$.

{}To illustrate the above discussion we note that the root lattices
of simply-laced Lie groups are even. The groups that have prime $N$
are the following: ({\em i})
$SU(n)$, $n$ is an odd prime, and $N=n$;
({\em ii}) $E_6$, $N=3$;
({\em iii}) SO(8n), $N=2$; ({\em iv}) $E_8$, $N=1$.

{}Similar considerations apply to right-moving chiral lattices, and also
Lorentzian lattices. In the latter case all the scalar products of vectors
are understood with respect to the Lorentzian signature.

%==============================================================================
\section{$N2$ Model}
\bigskip

{}Since the $N2$ model of subsection \ref{subsecIVB}
has $N=4$ space-time supersymmetry, it must correspond
to one of the four-dimensional $N=4$ supersymmetric models classified by
Narain \cite{Narain}, {\em i.e.}, there must be a choice of the constant
background fields such that the corresponding
even self-dual Lorentzian lattice is spanned by momenta (\ref{N2}). Here we
briefly construct such a lattice.

{}Consider a general lattice $\Gamma^{2,18}$ spanned by the vectors
\begin{equation}
 P=k^i m_i +k_i n^i +k^J p^J  ~,~~~m_i ,n^i \in {\bf Z}~,~~~
 p^J \in \Gamma^{16}~,
\end{equation}
where ($i=1,2$)
\begin{eqnarray}
 &&k^i =({1\over 2}{\ell}^{\ast i} ; {1\over 2}{\ell}^{\ast i} ;
 {\bf 0} )~,\\
 &&k_i =(-(B_{ji} +{1\over 4}A^I_j A^I_i){\ell}^{\ast j} -{\ell}_i ;
 -(B_{ji} +{1\over 4}A^I_j A^I_i){\ell}^{\ast j} +{\ell}_i ;A^I_i )~,\\
 &&k^J =(-{1\over 2}{\ell}^{\ast i} A^J_i ;
 -{1\over 2}{\ell}^{\ast i} A^J_i ; \delta^{IJ} )
\end{eqnarray}
are Lorentzian vectors of signature $((-)^2 ,(+)^{18})$, and
\begin{equation}
 {\ell}^{\ast i} \cdot {\ell}_j =\delta^i_j ~,~~~{\ell}_i
 \cdot {\ell}_j =g_{ij} ~,~~~{\ell}^{\ast i} \cdot
 {\ell}^{\ast j} =g^{ij} ~,
\end{equation}
$g_{ij}$, $B_{ij}$ and $A^I_i$ being constant background
symmetric, anti-symmetric and gauge (Wilson lines) fields, respectively.
Suppose that ${\ell}^{\ast i}=2{\zeta}^i$ and
${\ell}_j ={1\over 2} {\zeta}^{\ast}_i$, where $\{{\zeta}^i m_i\}
=\Gamma^2$ ($SU(3)$ root lattice), and $\{{\zeta}^{\ast}_i n^i \}=
{\tilde \Gamma}^2$ ($SU(3)$ weight lattice), and in the following we will use
the convention where ${\zeta}^1 \cdot {\zeta}^1 =
{\zeta}^2 \cdot {\zeta}^2 =-2{\zeta}^1 \cdot {\zeta}^2 =2$.
Then, provided that $A^I_1 = -A^I_2 \equiv A^I$, $2B_{12} =-2B_{21}\in {\bf Z}+
{1\over2}$, and ${1\over 2}(A^I)^2 \in{\bf Z}+{2\over 3}$, the momenta $P$
can be expressed as
\begin{equation}
 P=({\overline p}; {p} ; p^I +\alpha_2 A^I )~,
\end{equation}
where ${\overline p} \in 3A^I p^I {w} +\Gamma^2$, ${p}
\in (3A^I p^I +\alpha_2) {w}
+\Gamma^2$ and $p^I \in \Gamma^{16}$ (Here $\alpha_2 \equiv n^1 -n^2$, and
${w}\equiv {\zeta}^{\ast}_2 -{\zeta}^{\ast}_1$).

{}The lattice $\Gamma^{2,2} \otimes \Gamma^{2,2}
\otimes \Gamma^{2,18}$ has exactly
the momentum spectrum of the model $N2$.
This proves that the latter does describe an $N=4$ space-time
supersymmetric heterotic string model with a Wilson line.

%============================================================================
%\newpage

\end{document}